\pdfoutput=1
\documentclass[sigconf]{acmart}
\usepackage[ruled,linesnumbered]{algorithm2e}
\usepackage{algpseudocode}
\usepackage{multirow}
\usepackage{amsmath}
\usepackage{algpseudocode}
\usepackage{pifont}
\usepackage{enumerate}
\usepackage{bm}
\usepackage{balance}
\newtheorem{property}{Property}
\newcommand{\Rmnum}[1]{\uppercase\expandafter{\romannumeral #1\relax}}
\AtBeginDocument{%
  \providecommand\BibTeX{{%
    \normalfont B\kern-0.5em{\scshape i\kern-0.25em b}\kern-0.8em\TeX}}}

\copyrightyear{2023} 
\acmYear{2023} 
\setcopyright{acmlicensed}\acmConference[MM '23]{Proceedings of the 31st
ACM International Conference on Multimedia}{October 29-November 3,
2023}{Ottawa, ON, Canada}
\acmBooktitle{Proceedings of the 31st ACM International Conference on
Multimedia (MM '23), October 29-November 3, 2023, Ottawa, ON, Canada}
\acmPrice{15.00}
\acmDOI{10.1145/3581783.3611920}
\acmISBN{979-8-4007-0108-5/23/10}

\acmSubmissionID{1004}

\begin{document}
\title{Securing Fixed Neural Network Steganography}

\author{Zicong Luo}
\email{zcluo21@m.fudan.edu.cn}
\affiliation{%
	\institution{Fudan University}
	\streetaddress{Handan Road 220}
	\city{Shanghai}
	\country{China}}

\author{Sheng Li}
\authornote{Corresponding author: Sheng Li}
\email{lisheng@fudan.edu.cn}
\affiliation{%
	\institution{Fudan University}
	\streetaddress{Handan Road 220}
	\city{Shanghai}
	\country{China}}

\author{Guobiao Li}
\email{20210240200@fudan.edu.cn}
\affiliation{%
	\institution{Fudan University}
	\streetaddress{Handan Road 220}
	\city{Shanghai}
	\country{China}}
 
\author{Zhenxing Qian}
\email{zxqian@fudan.edu.cn}
\affiliation{%
	\institution{Fudan University}
	\streetaddress{Handan Road 220}
	\city{Shanghai}
	\country{China}}
 
\author{Xinpeng Zhang}
\email{zhangxinpeng@fudan.edu.cn}
\affiliation{%
	\institution{Fudan University}
	\streetaddress{Handan Road 220}
	\city{Shanghai}
	\country{China}}

\renewcommand{\shortauthors}{Zicong Luo, Sheng Li, Guobiao Li, Zhenxing Qian, \& Xinpeng Zhang}

\begin{abstract}
Image steganography is the art of concealing secret information in images in a way that is imperceptible to unauthorized parties. Recent advances show that is possible to use a fixed neural network (FNN) for secret embedding and extraction. Such fixed neural network steganography (FNNS) achieves high steganographic performance without training the networks, which could be more useful in real-world applications. However, the existing FNNS schemes are vulnerable in the sense that anyone can extract the secret from the stego-image. To deal with this issue, we propose a key-based FNNS scheme to improve the security of the FNNS, where we generate key-controlled perturbations from the FNN for data embedding. As such, only the receiver who possesses the key is able to correctly extract the secret from the stego-image using the FNN. In order to improve the visual quality and undetectability of the stego-image, we further propose an adaptive perturbation optimization strategy by taking the perturbation cost into account. Experimental results show that our proposed scheme is capable of preventing unauthorized secret extraction from the stego-images. Furthermore, our scheme is able to generate stego-images with higher visual quality than the state-of-the-art FNNS scheme, especially when the FNN is a neural network for ordinary learning tasks. 
\end{abstract}

\ccsdesc[500]{Security and privacy~Security services}
\ccsdesc[500]{Information systems~Multimedia information systems}

\keywords{Steganography, Fixed neural network, Key-controlled perturbation}



\maketitle

\section{INTRODUCTION}
The purpose of image steganography is to hide the secret imperceptibly in a cover image for covert communication, where only the receiver can accurately extract the secret from the stego-image (i.e., the image with a hidden secret)\cite{anderson1998limits}. The stego-image must be indistinguishable from the cover image visually and statistically to avoid being detected. 

Earlier steganographic methods hide the secret by modifying the least significant bits of the pixels in the cover image. Later, researchers follow the syndrome-trellis codes (STCs) steganographic framework\cite{filler2011minimizing} to minimize the distortion caused due to data hiding. A vast amount of functions have been proposed to measure such distortion for STC-based image steganography, including WOW\cite{holub2012designing}, S\_UNIWARD\cite{holub2013digital}, HILL\cite{li2014new}, and so on.

Recently, deep neural network (DNN)-based steganography has significantly altered the steganographic field thanks to the power of deep learning techniques. DNN-based steganography transforms the handcrafted conventional steganography into a data-driven and learning-based approach \cite{zhu2018hidden,baluja2017hiding,hu2018novel}. A typical DNN-based steganography contains a secret encoder network to embed the secret into a cover image, and a secret decoder network to extract the secret from a stego-image. These two networks have to be jointly or separately learnt to achieve high undetectability and data extraction accuracy. It requires a large amount of data and computational resources to train good steganographic networks (i.e., the secret encoder or decoder network). On the other hand, the steganographic networks are relatively large in size compared with conventional steganographic tools. It usually requires more than 100MB to store a secret encoder or decoder network. It raises our concerns regarding how we could covertly transmit the steganographic networks to the sender and receiver who may not possess any steganographic tools. 

\begin{figure*}[htbp]
	\centering
	\includegraphics[width=0.88\textwidth]{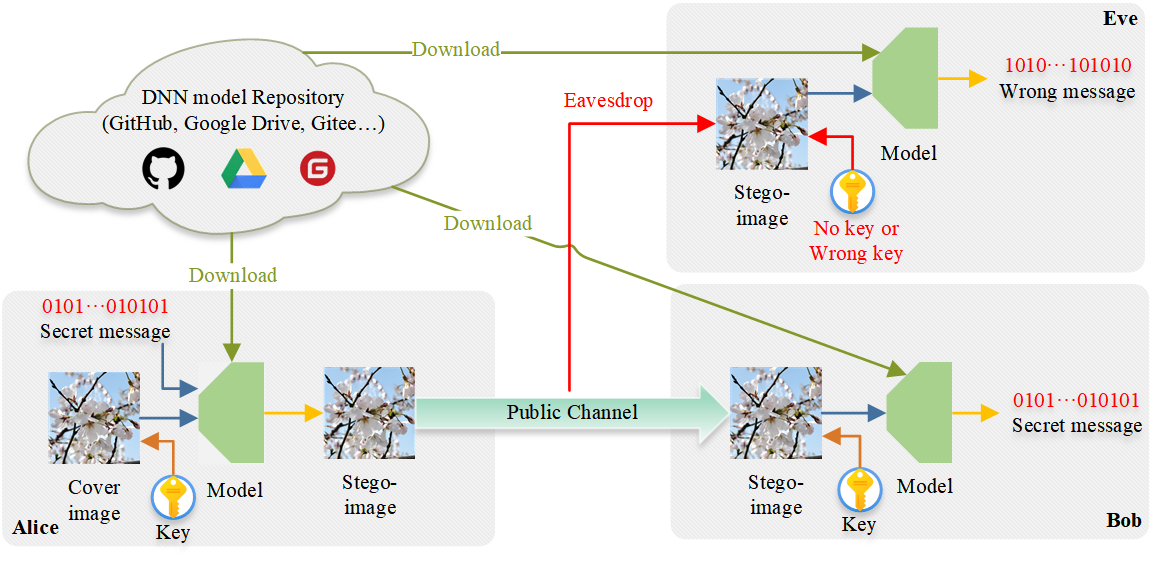}
	\caption{The application scenario of our proposed method. Alice (sender) and Bob (receiver) download a FNN from a DNN model repository. Alice uses this FNN to generate a stego-image with a key and sent it to Bob through a public communication channel. Bob can extract the secret from the stego-image using the FNN with the corresponding key. Eve (attacker) may eavesdrop on the channel to intercept the stego-image. However, he is not able to extract the secret without knowing the key.}
	\label{fig_bg}
\end{figure*}

To avoid training the steganographic networks, researchers propose Fixed Neural Network Steganography (FNNS) \cite{kishore2021fixed,ghamizi2021evasion} to perform data embedding and extraction using a fixed neural network (FNN). Such kind of schemes keeps the neural network parameters fixed and modifies the cover image in a way similar to adversarial perturbation so that the FNN can decode the secret from the stego-image. It does not require network training and alleviates the problem of transmitting steganographic networks. However, if the FNN is exposed or a similar network is trained by an attacker, he can easily extract the secret from the stego-image. To address this issue, an intuitive approach is to encrypt the secret before data embedding. The problem is that the DNN-based steganographic schemes could not guarantee lossless data extraction due to the uncertainty of the neural networks. A single bit of extraction error of the cipher text would cause a failure in decryption. 

In this paper, we propose a key-based FNNS to improve the security of the existing FNNS schemes. Instead of directly encrypting the secret, we propose to use a key to control the generation of the adversarial perturbations for data embedding using a FNN. Once the stego-image is generated, only the receiver who possesses the correct key is able to perform correct secret decoding, as shown in Fig.\ref{fig_bg}. To improve the visual quality and undetectability of the stego-images, we propose to estimate the perturbation cost and incorporate it into the design of the loss function to adaptively learn the perturbation from the FNN for data embedding. In particular, pixels with high perturbation costs will be assigned with low perturbation strength. Experimental results demonstrate the advantage of our scheme for preventing unauthorized data extraction. Furthermore, our scheme offers higher visual quality and undetectability than the state-of-the-art FNNS scheme, especially when using FNNs which work on ordinary learning tasks. 

The main contributions of this paper are summarized as follows.
\begin{enumerate}[1)]
  \item We are the first to look into the vulnerability of the existing FNNS schemes and propose a key-based FNNS to prevent unauthorized secret extraction from the stego-image.  
  \item We propose a key-based perturbation generation strategy by encrypting the stego-image before feeding it into the FNN for secret decoding. 
  
  \item We propose to estimate the perturbation cost of each image pixel, which is incorporated into the loss function to generate adaptive perturbation for data embedding. This is shown to be able to significantly improve the visual quality and undetectability of the stego-images.
\end{enumerate}

\section{RELATED WORKS}
\subsection{Traditional image steganography}
Traditional image steganography designs hand-crafted schemes to modify the cover image for data embedding, which can be divided into two categories including spatial domain-based steganography \cite{chan2004hiding,mielikainen2006lsb,li2009generalization,van1994digital,tsai2009reversible} and transform domain-based steganography\cite{westfeld2001f5,meerwald2001digital,mali2012robust,wang2013high,tao2018towards}. The former directly alters the pixel values in the spatial domain, while the latter changes the coefficients of the cover image in the transform domain to accommodate the secret. 

To improve the undetectability of the stego-images, researchers propose adaptive image steganography which can be applied to perform data embedding in the spatial or transformed domain\cite{pevny2010using,holub2012designing,holub2013digital,li2014new}. The most popular framework for adaptive steganography is the syndrome-trellis codes (STCs) steganographic framework \cite{filler2011minimizing}, which is able to achieve minimum distortion caused by data embedding. To effectively estimate the distortion, Pevn{\`y} \emph{et al.} \cite{pevny2010using} propose HUGO to measure the distortion for spatial domain-based steganography. Li \emph{et al.} \cite{li2014new} introduce HILL which exploits both the high-pass filter and low-pass filter to focus more on the texture area for data embedding. Kin-Cleaves \emph{et al.}\cite{kin2018adaptive} propose Dual-Syndrome Trellis Codes (Dual-STCs) to improve the robustness of steganography. On top of this, Guan \emph{et al.}\cite{guan2022double} propose a novel coding scheme that extends Dual-STCs to a double-layered embedding scheme which leverages the channel knowledge for data embedding. The capacity of these schemes is usually limited to ensure high undetectability. 

\subsection{DNN-based image steganography}
DNN-based image steganography trains a secret encoder for embedding secret into a cover image and a secret decoder for data extraction from a stego-image, which is shown to be promising to improve the performance of steganography. 

Zhu \emph{et al.}\cite{zhu2018hidden} pioneer the research for DNN-based image stegano-graphy, where an end-to-end autoencoder is proposed for data embedding. This is further improved by SteganoGAN \cite{zhang2019steganogan} which is able to achieve a payload of up to 6 bits per pixel (BPP). Tancik \emph{et al.}\cite{tancik2020stegastamp} incorporate the image printing and recapturing process in the encoder-decoder to enhance the performance of the secret decoder, which is robust against the attacks caused due to printing and recapturing. Baluja\cite{baluja2017hiding} proposes a DNN which is able to hide a color image into another color image. Wei \emph{et al.}\cite{wei2022generative} utilize generative adversarial networks (GANs)\cite{goodfellow2014generative} to directly generate stego-images from secrets without using a cover image.  Recently, researchers attempt to conduct data embedding using invertible networks \cite{jing2021hinet,xu2022robust,lu2021large,guan2022deepmih}, which treat the data embedding and extraction as a pair of inverse problems to achieve a high data embedding capacity. 

These schemes require training the steganographic networks on a large dataset. To avoid training, a few studies have been explored for Fixed Neural Network Steganography (FNNS) which does not require any training for data embedding and extraction. This is achieved by adding adversarial perturbations into a cover image to generate a stego-image which is able to produce some specific outputs corresponding to the secret. Ghamizi \emph{et al.}\cite{ghamizi2021evasion} produce the stego-images by encoding the secret into image labels to generate the perturbation, the capacity of which is rather limited. Kishore \emph{et al.}\cite{kishore2021fixed} propose to generate perturbations according to a message loss to produce the stego-image, which significantly improves the data embedding capacity compared with the work in \cite{ghamizi2021evasion}. 

Despite the advantage, the existing FNNS schemes are weak in securing the secret embedded in the stego-image. The attackers can extract the secret from the stego-images using the FNN or a surrogate network. On the other hand, it is yet unanswered on how we could generate a piece of perturbation that is able to minimize the distortion caused by data embedding. To address these two issues, we propose in this paper to generate key controlled and adaptive adversarial perturbations for FNNS. The former makes sure that the secret can only be extracted from the stego-image using a correct key, while the latter adaptively changes the perturbation strength for different pixels to improve the visual quality and undetectability of the stego-image.

\begin{figure*}[htbp]
	\centering
	\includegraphics[width=0.85\textwidth]{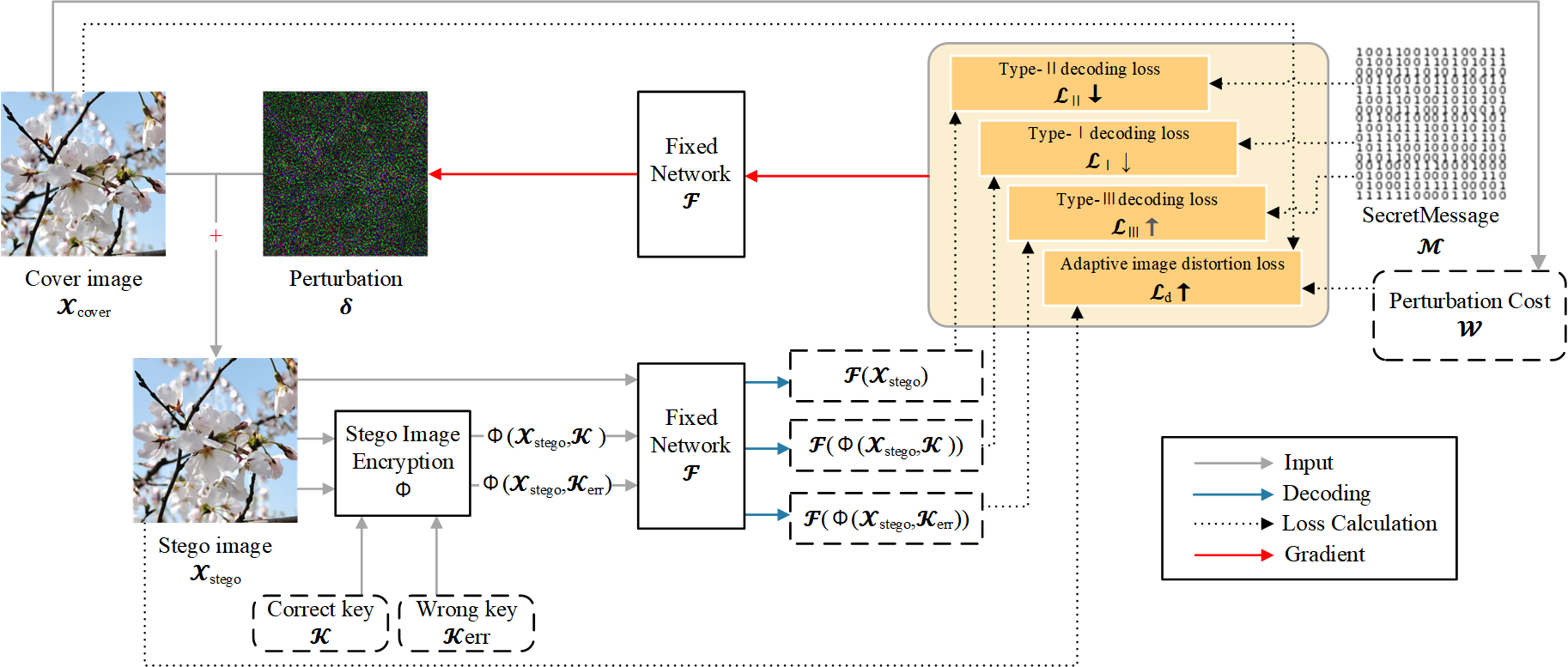}
	\caption{An overview of the proposed key-based FNNS.}
	\label{fig_framework}
\end{figure*}

\section{Problem Formulation}   \label{sec_Formulation}
Given a FNN $\mathcal{F}$, a cover image $\mathcal{X}_{cover}$, a secret $\mathcal{M}$ and different keys $\mathcal{K}$ and $\mathcal{K}_{err}$, our goal is to generate a stego-image by
\[
\mathcal{X}_{stego}=\mathcal{X}_{cover}+\Delta(\mathcal{F}, \mathcal{X}_{cover}, \mathcal{K},\mathcal{M}, \Phi),
\]
where $\Phi$ is an image encryption operation, $\Delta$ refers to our key controlled and adaptive perturbation generation scheme. Tab. \ref{table_notation} summarizes the notations used in this paper. 

The stego-image should have the least distortion compared with the cover image, which has to satisfy the following properties. 

\begin{property}
We should be able to extract the secret from the stego-image using the FNN and the correct key, i.e.,
    \label{prop:1}
\end{property}
\begin{equation}
    \mathcal{F}(\Phi(\mathcal{X}_{stego},\mathcal{K})) = \mathcal{M}.
\end{equation}

\begin{property}
We should not be able to extract the secret from the stego-image by only using the FNN, i.e.,
    \label{prop:2}
\end{property}
\begin{equation}
    \mathcal{F}(\mathcal{X}_{stego}) \neq \mathcal{M}.
\end{equation}

\begin{property}
We should not be able to extract the secret from the stego-image by using the FNN and a wrong key, i.e.,
    \label{prop:3}
\end{property}
\begin{equation}
    \mathcal{F}(\Phi(\mathcal{X}_{stego},\mathcal{K}_{err})) \neq \mathcal{M}.
\end{equation}
\begin{table}[tbp]
\centering
\caption{Notation.}
\begin{tabular}{c|l}
\hline
Notation & Description \\ \hline
$\mathcal{X}_{cover}$ & the cover image \\ \hline
$\mathcal{X}_{stego}$ & the stego-image \\ \hline
$\mathcal{F}$ & the FNN \\ \hline
$\mathcal{M}$ & the secret message \\ \hline
$\Phi$ & the process of image encryption \\ \hline
$\mathcal{\delta}$ & the perturbation added to the cover image \\ \hline
$\mathcal{K}$ & the correct key \\ \hline
$\mathcal{K}_{err}$ & the wrong key \\ \hline
$\mathcal{W}$ & the perturbation cost matrix \\ \hline
\end{tabular}
\label{table_notation}
\end{table}

\section{THE PROPOSED METHOD}
The overall structure of our proposed method is illustrated in Fig.\ref{fig_framework}. Given an RGB cover image $\mathcal{X}_{cover}\in[0,1]^{C\times H\times W}$, where $C$, $H$, and $W$ refer to the channel, height, and width of the image, respectively. We propose to generate key controlled and adaptive perturbations using $\mathcal{F}$ based on $\mathcal{K}$ and $\mathcal{M}\in \{0,1\}^{D\times H\times W}$ with $D$ being the number of bits per pixel to be hidden. We propose to encrypt the stego-image using $\mathcal{K}$ before optimizing the perturbations. And we design three different types of decoding loss terms to satisfy the three requirements listed in the previous section. To achieve the minimum data embedding distortion, we propose an adaptive image distortion loss by taking the perturbation cost for each pixel (in $\mathcal{X}_{cover}$) into consideration. The adaptive image distortion loss and the three types of decoding losses are combined to generate $\mathcal{X}_{stego}$ by adaptively adding the perturbations into $\mathcal{X}_{cover}$.     

\subsection{Stego-image Encryption}
We adopt a simple and straightforward way to encrypt the stego-image by adding each element in the stego-image with a random number. The encrypted version of the stego-image is computed by 
\begin{equation}
    \mathcal{X}^{\mathcal{K}}_{stego}=\Phi(\mathcal{X}_{stego}, \mathcal{K})=\mathcal{X}_{stego} \oplus Rand(\mathcal{K}),
\end{equation}
where $\oplus$ is the element-wise addition, and $Rand$ is a function to generate a random matrix with elements ranging [-1,1] using $\mathcal{K}$. Note that the dimension of the random matrix should be the same as that of the stego-image. 

\subsection{Perturbation Cost Estimation}
To enhance the visual quality and undetectability of the stego-images, we adopt a similar distortion function which has been used in the existing adaptive image steganographic schemes to estimate a perturbation cost for each pixel in $\mathcal{X}_{cover}$. The distortion function we use here is motivated by HILL\cite{li2014new}, which employs a high-pass filter $F_{h}$ and two low-pass filters $F_{l}^{1},F_{l}^{2}$ to measure the distortion caused by data embedding. The distortion function is formulated below
\begin{equation}
\mathcal{W} = \frac{1}{\lvert\mathcal{X}_{cover}\otimes F_{h}\rvert \otimes F_{l}^{1}} \otimes F_{l}^{2},
\end{equation}
where $F_{l}^{1}$ and $F_{l}^{2}$ are average filters with the size of $3\times3$ and $15\times15$, $\bigotimes$ refers to the convolution operation, and $F_h$ is designed as 
\begin{equation}
F_{h} = 
\begin{bmatrix}
-1 & 2 & -1 \\
2 & -4 & 2\\
-1 & 2 & -1
\end{bmatrix}.
\end{equation}

To make the perturbation cost suitable for optimizing the adaptive perturbations, $\mathcal{W}$ is further truncated and processed by 
\begin{equation}
\mathcal{W} = \begin{cases}
T & \text{if } \mathcal{W} > t \\
\mathcal{W} & \text{otherwise}
\end{cases}.
\end{equation}
The perturbation cost $\mathcal{W}$ measures the cost of perturbing each pixel in the cover image. Pixels with high perturbation costs are not suitable to be changed for data embedding, which refers to pixels' smooth areas. During the optimization, the perturbation strength should be low on such pixels. On the contrary, we can carry out perturbations with high strength for pixels with low perturbation costs.     

\subsection{Loss Functions}
We design four loss functions including adaptive image distortion loss, Type-\Rmnum{1} decoding loss, Type-\Rmnum{2} decoding loss, and Type-\Rmnum{3} decoding loss. The adaptive image distortion loss aims to preserve the quality of the stego-image according to the perturbation cost. The three decoding losses are used to guarantee the generation of key control perturbations.

\textbf{Adaptive image distortion loss.}
The adaptive image distortion loss is formulated by
\begin{equation}
\mathcal{L}_{d}(\mathcal{X}_{cover},\mathcal{X}_{stego},\mathcal{W}) =  \sqrt{\mathcal{W} \cdot(\mathcal{X}_{cover} - \mathcal{X}_{stego})^2} ,
\end{equation}
where $\cdot$ denotes the element-wise product operation. This is a weighted L2 distance between the cover image and the stego-image. By using such a loss, we shall be able to learn an adaptive perturbation during optimization, which focuses more on the pixels with low perturbation costs.  

\textbf{Type-\Rmnum{1} decoding loss}
We design a Type-\Rmnum{1} decoding loss below to satisfy Property \ref{prop:1} mentioned in Section \ref{sec_Formulation}.
\begin{equation}
\mathcal{L}_{\Rmnum{1}}(\mathcal{X}_{stego},\mathcal{K},\mathcal{M}) =  L_{BCE}(\mathcal{F}(\Phi(\mathcal{X}_{stego},\mathcal{K})),\mathcal{M}), 
\end{equation}
where $L_{BCE}$ a binary cross-entropy. This loss makes sure that, when the correct key is used to encrypt the stego-images, we are able to recover the secret by feeding the encrypted version of the stego-image into the FNN. 

\textbf{Type-\Rmnum{2} decoding loss.}
We design a Type-\Rmnum{2} decoding loss below to satisfy Property \ref{prop:2} mentioned in Section \ref{sec_Formulation}.
\begin{equation}
\mathcal{L}_{\Rmnum{2}}(\mathcal{X}_{stego},\mathcal{M}) =  L_{BCE}(\mathcal{F}(\mathcal{X}_{stego}),\mathcal{M}).
\end{equation}
This loss is used to make the decoded message dissimilar to the secret when no key is used for message decoding, i.e., we directly input the stego-image into the FNN to extract the secret.  

\textbf{Type-\Rmnum{3} decoding loss.}
We design a Type-\Rmnum{3} decoding loss below to satisfy Property \ref{prop:3} mentioned in Section \ref{sec_Formulation}.
\begin{equation}
\mathcal{L}_{\Rmnum{3}}(\mathcal{X}_{stego},\mathcal{K}_{err},\mathcal{M}) =  L_{BCE}(\mathcal{F}(\Phi(\mathcal{X}_{stego},\mathcal{K}_{err})),\mathcal{M}).
\end{equation}
This loss is used to prevent the attacker from using the wrong key for message decoding. To well simulate the scenario in which the user uses several randomly guessed keys for message decoding, we generate a wrong key set $\mathcal{K}_{err}$ containing $N$ distinct keys which are different from $\mathcal{K}$. We accumulate the Type-\Rmnum{3} decoding loss for each wrong key by    
\begin{equation}
\mathcal{L}_{\Rmnum{3}}(\mathcal{X}_{stego},\mathcal{K}_{err},\mathcal{M}) =  \sum_{n=1}^NL_{BCE}(\mathcal{F}(\Phi(\mathcal{X}_{stego},K_{err}^n)),\mathcal{M}),
\end{equation}
where $K_{err}^n$ denotes the $n^{th}$ key in $\mathcal{K}_{err}$.  

\textbf{Total Loss.}
The overall loss function is a weighted sum among the aforementioned losses, which is given by
\begin{equation}
\mathcal{L}_{Total} = \lambda_d  \mathcal{L}_{d} + \lambda_{\Rmnum{1}}  \mathcal{L}_{\Rmnum{1}} - \lambda_{\Rmnum{2}}  \mathcal{L}_{\Rmnum{2}} - \lambda_{\Rmnum{3}} \mathcal{L}_{\Rmnum{3}},
\end{equation}
where $\lambda_d$, $\lambda_{\Rmnum{1}}$, $\lambda_{\Rmnum{2}}$ and $\lambda_{\Rmnum{3}}$ are the weights for balancing different loss terms.

\subsection{Optimization strategy}
Our goal is to optimize the following problem:
\begin{equation}
\begin{split}
& min\quad \mathcal{L}_{Total} ,\\
& s.t.\quad 0 \leq \mathcal{X}_{stego} \leq 1.
\end{split}
\end{equation}

Researchers have proposed a lot of optimization approaches to solve this problem by generating adversarial perturbations. Note that we have to quantize the stego-images for real-world applications. The quantization will reduce the visual quality of the stego-image. It would also make the output of the FNN cross the decision boundary, which causes a reduction in message decoding accuracy. We propose a two-stage optimization strategy to alleviate the negative impact of quantization on steganographic performance. In the first stage, we carry out the optimization by only using the adaptive image distortion loss and Type-\Rmnum{1} decoding loss. In the second stage, we conduct the optimization of the total loss. As such, the output of the network would repeatedly cross the decision boundary during the optimization process to achieve stable performance. 

The details of the optimization process are given in Algorithm \ref{algo:stego}, where we use L-BFGS\cite{fletcher2013practical} as the main optimization approach. For each iteration, we quantize each element in each of the RGB channels of the stego-image into 255 levels to learn perturbations that are effective on quantized RGB images.

\begin{algorithm}[t]
	\caption{Key-based FNNS}
	\label{algo:stego}
	\KwIn{$\mathcal{F}$; $\mathcal{X}_{cover}$; $\mathcal{W}$;
$\mathcal{M}$; $\mathcal{K}$; $\mathcal{K}_{err}$}
	\KwOut{$\mathcal{X}_{stego}$}  
    \textbf{Parameters}: $\alpha$: learning rate; $E$: number of iterations; $st_1$: the number of steps for the first stage ; $st_2$:  the number of steps for the second stage;
    $\lambda_{d}, \lambda_{\Rmnum{1}}, \lambda_{\Rmnum{2}}, \lambda_{\Rmnum{3}}$;
    
	\BlankLine
 
	Freeze $\mathcal{F}$;  \Comment{Freeze the decoder network parameters}
 
	$\mathcal{X}_{stego} \leftarrow \mathcal{X}_{cover}$;   \Comment{Initialize}
 
	\For{$i = 1$ \KwTo $E$}{
        $\mathcal{L}_{Total}^1 \leftarrow \lambda_d  \mathcal{L}_{d} + \lambda_{\Rmnum{1}}  \mathcal{L}_{\Rmnum{1}}$ 
        
		$\mathcal{\delta} \leftarrow$ L-BFGS($\mathcal{L}_{Total}^1, \alpha, st_1$);  \Comment{Takes $st_1$ steps to optimize  $\mathcal{L}_{Total}^1$}

        $\mathcal{L}_{Total}^2 \leftarrow \lambda_d  \mathcal{L}_{d} + \lambda_{\Rmnum{1}}  \mathcal{L}_{\Rmnum{1}} - \lambda_{\Rmnum{2}}  \mathcal{L}_{\Rmnum{2}} - \lambda_{\Rmnum{3}}  \mathcal{L}_{\Rmnum{3}}$
        
		$\mathcal{\delta} \leftarrow$ L-BFGS($\mathcal{L}_{Total}^2,\alpha, st_2$);   \Comment{Takes $st_2$ steps to optimize $\mathcal{L}_{Total}^2$}
  
		$\mathcal{X}_{stego} \leftarrow clip_{1}^{0}(\mathcal{X}_{cover}+\mathcal{\delta})$;  \Comment{Clip the image to [0,1]}
		
		$\mathcal{X}_{stego} \leftarrow quantize(\mathcal{X}_{stego})$;    \Comment{Quantization}
	}
return $\mathcal{X}_{stego}$
\end{algorithm}

\section{Experiments}
\subsection{Setup}
\textbf{Datasets and DNN models.}
To evaluate the effectiveness of our proposed method, we conduct experiments on three different data-sets, namely MS-COCO\cite{lin2014microsoft}, Div2k\cite{agustsson2017ntire}, and CelebA\cite{liu2015deep}. To prepare the data for our experiments, we follow different procedures for different datasets. For Div2k, which is a high-quality dataset consisting of diverse images, we use the entire validation set as our cover images. For MS-COCO and CelebA, which are large-scale datasets for natural scenes and human faces, we randomly use 100 images as our cover images. For all the datasets, we resize the images to a fixed resolution of $256\times256$ to ensure consistency and efficiency. To generate the secret and the key, we use a random function to assign each bit with equal probability. Specifically, we generate each bit from a Bernoulli distribution with a parameter of 0.5, meaning that each bit has a 50\% chance of being 0 or 1. As such, the secret and the key are uniformly distributed and independent. We use a SteganoGAN\cite{zhang2019steganogan} model pre-trained on the corresponding dataset as the FNN. The SteganoGAN is a popular tool for image steganography which is designed using generative adversarial networks.

\textbf{Parameters and Evaluation Metrics.}
We set the learning rate $\alpha$ as 0.10, the number of iterations $E$ as 100, and the number of the two-stage L-BFGS optimizations as 15 for both $st_1$ and $st_2$. The parameters for balancing different loss terms are set as $\lambda_{d}=40$, $\lambda_{\Rmnum{1}}=5$, $\lambda_{\Rmnum{2}}=0.05$ and $\lambda_{\Rmnum{3}}=0.05$, respectively. The number of wrong keys is set as $N=3$ in the wrong key set $\mathcal{K}_{err}$. The $t$ and $T$ for computing the perturbation cost are set as 0.5 and 3, respectively. We use three widely used metrics to evaluate the performance of our method, including bit error rate (BER), peak signal-to-noise ratio (PSNR), and structural similarity index (SSIM). The BER measures the accuracy of the secret extraction, while the PSNR and SSIM measure the visual quality of stego-images.

\subsection{Comparisons}

\begin{table*}[htb]
\renewcommand{\arraystretch}{0.98} 
\caption{Performance comparisons among different methods on different datasets. $\uparrow$ indicates better, and vice versa. We mark the best-performing values in bold.}
\centering
\resizebox{1\textwidth}{!}{
\begin{tabular}{c|c|cccc|cccc|cccc}
    \hline
    \hline
    \multirow{2}{*}{Datasets} & \multirow{2}{*}{Methods} &\multicolumn{4}{c|}{BER(\%)$\downarrow$}   & \multicolumn{4}{c|}{PSNR(dB)$\uparrow$} & \multicolumn{4}{c}{SSIM$\uparrow$} \\ 
    \cline{3-14}
    &&1BPP&2BPP&3BPP&4BPP
    &1BPP&2BPP&3BPP&4BPP
    &1BPP&2BPP&3BPP&4BPP\\
    \hline
    \multirow{3}{*}{COCO} & SteganoGAN 
    &3.40&6.29&11.13&15.70
    &25.32&24.27&25.01&24.94
    &0.84&0.82&0.82&0.82 \\
    &FNNS
    &0.03&0.01&\textbf{0.01}&14.56
    &37.58&36.04&26.31&\textbf{34.75}
    &0.91&0.93&0.71&\textbf{0.91} \\
    &Ours
    &\textbf{2E-04}&\textbf{0.01}&0.10&\textbf{14.43}
    &\textbf{40.46}&\textbf{36.62}&\textbf{29.43}&33.65
    &\textbf{0.98}&\textbf{0.95}&\textbf{0.84}&0.89 \\
    \hline
    \multirow{3}{*}{Div2k} & SteganoGAN
    &5.12&8.31&13.74&22.85
    &21.33&21.06&21.42&21.84
    &0.76&0.76&0.77&0.78 \\
    &FNNS
    &\textbf{0.00}&\textbf{5E-04}&\textbf{0.07}&\textbf{2.21}
    &25.96&21.41&18.68&18.68
    &0.79&0.60&0.38&0.38    \\
    &Ours
    &0.01&0.08&1.87&8.77
    &\textbf{33.92}&\textbf{27.60}&\textbf{25.77}&\textbf{25.79}
    &\textbf{0.96}&\textbf{0.88}&\textbf{0.79}&\textbf{0.77} \\
    \hline
    \multirow{3}{*}{CelebA} & SteganoGAN
    &3.94&7.36&8.84&10.00
    &25.98&25.53&25.70&25.08
    &0.85&0.86&0.85&0.82 \\
    &FNNS
    &\textbf{2E-06}&\textbf{5E-05}&\textbf{4E-04}&\textbf{2.40}
    &34.43&34.48&30.98&30.79
    &0.83&0.87&0.80&0.75 \\
    &Ours
    &3E-04&2E-03&0.02&2.75
    &\textbf{39.48}&\textbf{36.43}&\textbf{33.22}&\textbf{33.86}
    &\textbf{0.95}&\textbf{0.92}&\textbf{0.87}&\textbf{0.86} \\
    \hline
\end{tabular}}
\label{table_comparison}
\end{table*}

\begin{table*}[htb]
    \centering
    \caption{The BER of the secret extracted using our method on different datasets without using a key or using wrong keys at different payloads.}
    \resizebox{.6\textwidth}{!}{
    \begin{tabular}{c|cccc|cccc}
    \hline
    \hline
    \multirow{2}{*}{Datasets} &\multicolumn{4}{c|}{BER without key(\%)$\uparrow$} & \multicolumn{4}{c}{BER with wrong key(\%)$\uparrow$} \\ 
    \cline{2-9}
    &1BPP&2BPP&3BPP&4BPP
    &1BPP&2BPP&3BPP&4BPP \\
    \hline
    COCO
    &29.77&24.81&13.64&28.98
    &34.32&30.51&18.86&32.62    \\
    Div2k
    &22.26&21.11&13.72&22.31
    &25.54&24.95&17.73&25.19    \\
    CelebA 
    &30.65&31.61&29.53&27.98
    &33.76&35.19&33.64&31.97    \\
    \hline
    \end{tabular}}
    \label{tab:err}
\end{table*}

We conduct quantitative comparisons between our proposed method and the state-of-the-art FNNS scheme proposed in \cite{kishore2021fixed} (termed as the FNNS for short). We follow the default settings according to the source code of FNNS for implementation, where we use a pre-trained SteganoGAN model \cite{zhang2019steganogan} as the FNN. Tab.\ref{table_comparison} reports the performance among different schemes in terms of BER, PSNR, and SSIM, where we use the same key for secret embedding and extraction using our proposed method. Compared with the FNNS, our scheme achieves similar BER which is close to zero when the payload is less than 3BPP. Our method outperforms the FNNS in terms of PSNR and SSIM in almost all cases. In Div2k, the PSNR and SSIM of our proposed method are significantly higher than the FNNS with 6dB and 0.2 increment in PSNR and SSIM, respectively. This indicates the effectiveness of our adaptive perturbation learning strategy by incorporating the perturbation cost during the optimization. We also observe that both our proposed method and the FNNS perform better than the pre-trained SteganoGAN, which indicates the advantage of the FNNS-based schemes. 

Tab.\ref{tab:err} further gives the performance of our proposed scheme when the attackers do not know the correct keys. We can see that the BER is around 30\% (50\% for random guess) for the no key and wrong key cases at different payloads and datasets. This is to say, the attackers are not able to correctly extract the secret from our stego-images if they only know the FNN. 

\begin{figure*}[htb]
\centering
  \includegraphics[width=.77\linewidth]{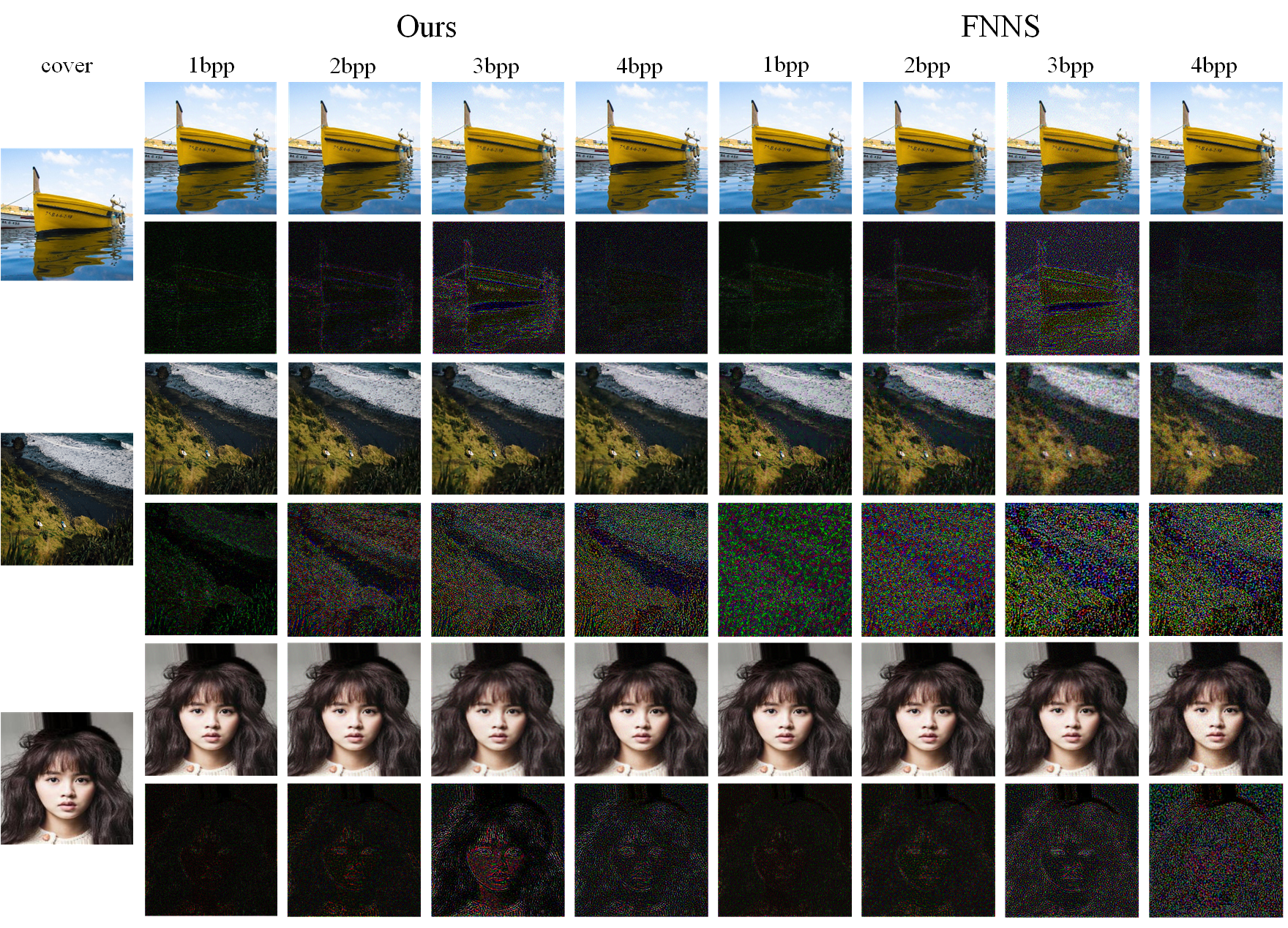}
  \caption{Illustration of stego-images generated using our method and the FNNS at different payload settings on different datasets. The first column shows the cover image, the second to the ninth column shows the stego-images generated using our method and the FNNS from a payload of 1BPP to 4BPP. For each cover image, we present the stego-images in the first row and the corresponding residuals (i.e., the absolute difference between the cover and stego-images) in the second row. We magnify the residuals by 10 times to highlight the altered regions. The cover images (from top to bottom) used in this figure are randomly selected from MS-COCO\cite{lin2014microsoft}, Div2k\cite{agustsson2017ntire}, and CelebA\cite{liu2015deep}, respectively.}
  \label{fig_qualitative}
\end{figure*}

Fig. \ref{fig_qualitative} illustrates some stego-images generated using our proposed method and the FNNS. It can be seen that our stego-images are visually similar to the corresponding cover images. Compared with the FNNS, our method is able to adaptively adjust the perturbation strength according to the image content, where more perturbations are learnt for texture areas. This further demonstrates the advantage of our proposed scheme over the FNNS in terms of visual quality.  
 
\subsection{Performance on non-steganographic DNN Models}
In this section, we evaluate the performance of our proposed method on DNN models which perform ordinary learning tasks. Concretely, we download two non-steganographic DNN models from a public model repository, including DnCNN\cite{zhang2017beyond} and FFDNet\cite{zhang2018ffdnet}, which are pre-trained for image denoising. We use these two image-denoising DNN models as the FNNs for image steganography.  
Tab.\ref{table_public} gives the comparison between our proposed method and the FNNS on these two DNN models. It can be seen that our scheme performs significantly better than the FNNS in terms of PSNR and SSIM. For both DNN models, the PSNR and SSIM of our stego-images are significantly higher than those generated using FNNS, with over 15dB improvement in PSNR and over 0.6 increment in SSIM. While the BER of our proposed method is still acceptable. Therefore, our proposed method would be more useful for real-world applications when the sender and receiver do not possess any steganographic DNN models. 

\subsection{Undetectability}
\begin{figure}[tbp]
\centering
  \includegraphics[width=.96\linewidth]{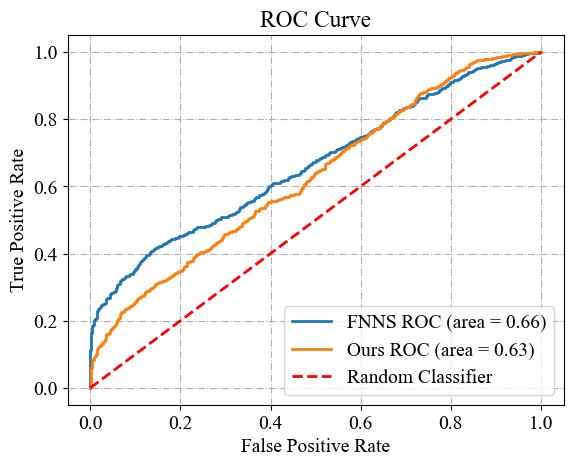}
  \caption{The ROC curves plotted for different schemes against the StegExpose.}
  \label{fig_ROC}
\end{figure}

One of the important criteria for measuring the steganographic performance is the undetectability of the stego-images against steganalysis tools, which are used to detect the existence of hidden secrets in an image. We follow the suggestion given in \cite{kishore2021fixed} to evaluate the undetectability, where StegExpose \cite{boehm2014stegexpose} is adopted as the steganalysis tool. To measure the undetectability of our method, we randomly select 1000 cover images from the MS-COCO dataset and generate 1000 stego-images using FNNS and our proposed method at a payload of 1 BPP. Fig.\ref{fig_ROC} plots the ROC curves for different schemes. It can be seen that the undetectability of our proposed method is better than that of the FNNS.  

\subsection{Ablation Study}
In this section, we conduct ablation studies to analyze the effects of different components in our method. In all the ablation studies, we use the CelebA dataset for evaluation and conduct a data embedding at a payload of 1BPP to generate the stego-images, where the Type-\Rmnum{1} decoding loss is always used.

\textbf{Effectiveness of perturbation cost.}
In order to evaluate the effectiveness of perturbation cost, we replace the adaptive image distortion loss with the standard L2 loss in our proposed scheme and re-evaluate the performance. The results are shown in the second row in Tab.\ref{tab:ablation}, where we report various metrics including the PSNR, SSIM, and the BER using different keys. Compared with the numbers shown in the last row in Tab.\ref{tab:ablation}, we can observe that using the perturbation cost improves the performance of BER, PSNR, and SSIM. Because it allows us to focus more on the low-cost regions corresponding to the texture areas for data embedding. 

\textbf{Effectiveness of Type-\Rmnum{2} decoding loss.}
To examine the effectiveness of the Type-\Rmnum{2} decoding loss, we take out this loss and rerun the experiments, the results of which are given in the third row in Tab.\ref{tab:ablation}. Compared with the results given in the last row in Tab.\ref{tab:ablation}, we can see the use of Type-\Rmnum{2} decoding loss would be able to make it more difficult for the attacker to decode the secret without using a correct key.

\textbf{Effectiveness of Type-\Rmnum{3} decoding loss.}
For the same token, we remove the Type-\Rmnum{3} decoding loss and rerun the experiments to conduct ablation studies for Type-\Rmnum{3} decoding loss. The corresponding results are reported in the fourth row in Tab.\ref{tab:ablation}. We see that using Type-\Rmnum{3} decoding loss substantially increases the BER for the cases where no key or wrong key is used for decoding. Compared with the results in the last row in Tab.\ref{tab:ablation}, using the Type-\Rmnum{3} decoding loss increase the BER of 8.29\% and 6.75\% for the no key and wrong key cases, respectively.

\begin{table*}[tb]
\renewcommand{\arraystretch}{1.0} 
\caption{Performance comparisons on the CelebA dataset using two non-steganographic DNN models, downloaded from a public model repository, as FNNs for steganography. }
\centering
\resizebox{1\textwidth}{!}{
\begin{tabular}{c|c|ccc|ccc|ccc|ccc|ccc}
    \hline
    \hline
    \multirow{2}{*}{Models} & \multirow{2}{*}{Methods} & \multicolumn{3}{c|}{BER(\%)$\downarrow$} & \multicolumn{3}{c|}{BER without key(\%)$\uparrow$} & \multicolumn{3}{c|}{BER with wrong key(\%)$\uparrow$}  & \multicolumn{3}{c|}{PSNR(dB)$\uparrow$} & \multicolumn{3}{c}{SSIM$\uparrow$} \\ 
    \cline{3-17}
    &&1BPP&2BPP&3BPP
    &1BPP&2BPP&3BPP
    &1BPP&2BPP&3BPP
    &1BPP&2BPP&3BPP
    &1BPP&2BPP&3BPP\\
    \hline
    \multirow{2}{*}{DnCNN} 
    &FNNS
    &\textbf{0.04}&\textbf{0.02}&\textbf{0.08}
    &/&/&/
    &/&/&/
    &13.64&10.38&11.76
    &0.12&0.07&0.07 \\
    &Ours
    &3.03&3.41&4.00
    &20.45&21.80&21.46
    &20.97&21.38&21.23
    &\textbf{31.81}&\textbf{30.39}&\textbf{29.19}
    &\textbf{0.80}&\textbf{0.75}&\textbf{0.70} \\
    \hline
    \multirow{2}{*}{FFDNet} 
    &FNNS
    &7.17&9.92&14.69
    &/&/&/
    &/&/&/
    &12.03&12.31&12.26
    &0.08&0.08&0.08 \\
    &Ours
    &\textbf{5.62}&\textbf{5.76}&\textbf{5.77}
    &21.28&22.40&21.82
    &21.72&22.14&21.95
    &\textbf{31.88}&\textbf{30.50}&\textbf{29.34}
    &\textbf{0.80}&\textbf{0.76}&\textbf{0.71} \\
    \hline
\end{tabular}}
\label{table_public}
\end{table*}

\begin{table*}[htb]
    \centering
    \caption{Effectiveness of different components in our method.}
    \resizebox{.9\textwidth}{!}{
    \begin{tabular}{cccc|ccccc}
    \hline
    \hline
    Perturbation  & Type-\Rmnum{2} & Type-\Rmnum{3} & Two-stage &\multirow{2}{*}{BER(\%)$\downarrow$}  & BER without & BER with & \multirow{2}{*}{PSNR(dB)$\uparrow$} & \multirow{2}{*}{SSIM$\uparrow$}  \\
    Cost&Loss&Loss&Update&& key(\%)$\uparrow$ & wrong key(\%)$\uparrow$ & \\
    \hline
    \ding{55} & \ding{55} & \ding{55} & \ding{55}  & 0.04 & 8.08 & 16.09 & 38.24 & 0.94  \\
    \ding{55} & \ding{51} & \ding{51} & \ding{51}  & 2E-03 & \textbf{30.91} & 33.67 & 38.58 & 0.94 \\
    \ding{51} & \ding{55} & \ding{51} & \ding{51}  & 4E-03 & 29.60 & 33.09 & 38.44 & 0.95  \\
    \ding{51} & \ding{51} & \ding{55} & \ding{51}  & 7E-03 & 22.35 & 27.01 & 39.16 & 0.95  \\
    \ding{51} & \ding{51} & \ding{51} & \ding{55}  & 0.04 & 25.38 & 28.74 & 37.11 & 0.93  \\
    \ding{51} & \ding{51} & \ding{51} & \ding{51}  & \textbf{3E-04} & 30.64 & \textbf{33.76} & \textbf{39.48} & \textbf{0.95}  \\
    \hline
    \end{tabular}}
    \label{tab:ablation}
\end{table*}

\begin{table*}[htb]
    \centering
    \caption{The effectiveness of iterative quantization.}
    \resizebox{.7\textwidth}{!}{
    \begin{tabular}{c|ccccc}
    \hline
    \hline
    &\multirow{2}{*}{BER(\%)$\downarrow$}  & BER without & BER with & \multirow{2}{*}{PSNR(dB)$\uparrow$} & \multirow{2}{*}{SSIM$\uparrow$}  \\
    && key(\%)$\uparrow$ & wrong key(\%)$\uparrow$ && \\
    \hline
    One-stage Update& 0.32 &33.58&36.34&38.19&0.95\\
    \hline
    Two-stage Update&0.65&35.56&38.13&37.97&0.95\\
    \hline
    \end{tabular}}
    \label{tab:quantize}
\end{table*}

\textbf{Effectiveness of two-stage update.}
Next, we conduct evaluations to see whether our two-stage update strategy is helpful. To do so, we use a single one-stage update by optimizing only the total loss using the L-BFGS optimizer. The results are shown in the fifth row in Tab.\ref{tab:ablation}. We find that our two-stage updating strategy indeed improves the steganographic performance because it helps us to escape local minima caused by quantization. 
Particularly, the BER is improved from 0.04\% to 3E-04\% by using our two-stage update strategy.

\textbf{Effectiveness of iterative quantization.}
To generate a stego-image that is suitable to be transmitted in public communication channels, we have to quantize the image after the optimization. A simple way is to quantize the image when the optimization is done. In our proposed method, we quantize the images in each iteration instead of performing the quantization in the final iteration. We believe such a strategy would be able to learn perturbations that are more appropriate for the quantized images. For justification, we conduct two additional experiments here, where we conduct one-stage updates and two-stage updates and only perform the image quantization after the optimization, respectively. Tab.\ref{tab:quantize} reports the results of these experiments. 
By comparing the results in Tab.\ref{tab:quantize} with the last two rows in Tab.\ref{tab:ablation}, we can clearly observe that the optimal performance is achieved only when both iterative quantization and two-stage updates are applied. 

In addition, as shown in the first row of Tab.\ref{tab:ablation}, the performance of our method degrades significantly when none of the proposed components are incorporated. We also observe that the combination of all components leads to a substantial improvement over the use of any single component. This indicates that our components are complementary for performance boosting.

\section{Conclusion}
In this paper, a key-based FNNS is proposed to improve the security of the existing FNNS schemes. Unlike the existing FNNS schemes, we use a key to control the generation of the adversarial perturbations for data embedding, which is performed by encrypting the input images of the FNN. Given a stego-image, only the receiver who possesses the correct key can extract the secret using the FNN.  We further propose an adaptive perturbation generation scheme by taking the perturbation cost into account during the optimization. This is shown to be effective in improving the visual quality and undetectability of the stego-images. We use a pre-trained steganographic network and two image-denoising DNN models as the FNNs to evaluate the performance of our key-based FNNS. The results indicate the advantage of our proposed scheme over the state-of-the-art FNNS in terms of preventing unauthorized secret extraction as well as the steganographic performance.    

\begin{acks}
This work is supported by National Natural Science Foundation of China under Grant 62072114, U20A20178, U20B2051, U1936214 and U22B2047.
\end{acks}

\bibliographystyle{ACM-Reference-Format}
\balance
\bibliography{ref}

\appendix

\end{document}